\documentclass[preprint]{aastex}
\usepackage{graphicx} 

\begin{document}

\title{Comparative Analysis of Peculiar Type Ia 1991bg-like Supernovae 
Spectra}
\author{Brandon A. Doull and E. Baron}
\affil{Homer L. Dodge Department of Physics and Astronomy,\\
		The University of Oklahoma,\\
		Norman, Oklahoma}
\email{doull@nhn.ou.edu}

\begin{abstract}
Spectroscopic analyses of Type Ia supernovae have shown there exist 
four spectroscopic groups---cools, broad line, shallow silicon, and core 
normal---defined by the widths of the Si~II features 
$\lambda5972$ and $\lambda6355$. 1991bg-likes are classified as ``cools". 
Cools are dim, undergo a rapid decline in luminosity, and produce 
significantly less $^{56}$Ni than normal Type Ia supernovae. They 
also have an unusually deep and wide trough in their spectra around 
4200~\AA\ and a relatively strong Si~II absorption attributed to $\lambda5972$. 
We examine the spectra of supernova (SN) 1991bg and the cools 
SN~1997cn, SN~1999by, and SN~2005bl using the highly parameterized 
synthetic spectrum code {\bf SYNOW}, and find general agreement with 
similar spectroscopic studies.  Our analysis reveals 
that this group of supernovae is fairly homogeneous, with many of 
the blue spectral features well fit by Fe~II. The nature of the 
spectroscopic commonalities and the variations in the class are 
discussed.  Finally, we examine intermediates such as SN~2004eo and 
discuss the spectroscopic subgroup distribution of Type~Ia supernovae.
\end{abstract}


\section{Introduction}
Type Ia supernovae (hereafter SNe~Ia) are important objects 
in the study of nucleosynthesis, stellar evolution, and 
modern cosmology. It is believed that SNe~Ia are thermonuclear 
explosions of carbon-oxygen white dwarfs which have approached 
to within about $1\%$ of the Chandrasekhar Mass 
(1.39$M_{\bigodot}$) \citep{hillebrandt00, hoeflich09}. 
In cosmology, SNe~Ia were suggested early on \citep{baade38} to 
be useful as standard candles for making accurate distance 
measurements in the determination of cosmological parameters, 
because of both their significant homogeneity in absolute bolometric magnitude 
and large apparent brightness. With modern supernova surveys such as the 
Supernova Cosmology Project, ESSENCE Supernova Survey, Nearby Supernova 
Factory, Palomar Transient Factory, Supernova Legacy Survey, and the Large 
Synoptic Survey Telescope, accurate measurements of these parameters 
based on SNe~Ia at varying redshifts have been and continue to be made 
\citep{perlmutter99, krisciunas08, snf06, ptf09, conley11, lsst09}.  
It has even been possible to measure the rate of 
expansion of the universe and put limits on cosmological models 
\citep{riess98, goobar00, woodvasey07, riess09, hicken09, chuang11, parkinson10}. 
It is because of this important role SNe~Ia play that a 
deep understanding of their properties, variations, and 
evolution is vital.

In our study we have narrowed our focus to the ``cool" spectroscopic 
subgroup, epitomized by SN~1991bg which occurred in NGC 4374, and so-called 
``intermediates" which could potentially bridge the gap between the four 
spectroscopic subgroups.   We used the highly-parameterized synthetic 
spectrum code {\bf SYNOW} to model the spectra of several cool SNe~Ia. 
To better understand the {\bf SYNOW} code and its use
see \cite{branch07, branch03, branch02}, and for more in depth technical 
details see \cite{fisher00}. The spectra range from a few days before 
{\emph B} maximum to as late as one month post {\emph B} maximum in the 
optical spectra with wavelengths from the Ca~II H\&K feature in the blue 
to the Ca~II infrared triplet. The spectra have been corrected for 
redshift of their host galaxies and normalized to remove the slope of 
the continuum, as we are primarily interested in spectral features 
for making line identifications.  This normalization, described in 
\cite{jeffery07}, also serves to facilitate the comparisons between different 
supernovae. Extremely noisy spectra have been smoothed using boxcar
smoothing. 

In \S\ref{sec:subgroups} the method of spectroscopic sub-classification 
is discussed. In \S\ref{sec:cools} the properties, progenitors, and 
possible explosion mechanisms of 
1991bg-likes are reviewed, and in \S\ref{sec:id} our line identifications are 
given and compared with similar studies. In \S\ref{sec:ints} we discuss 
the potential of intermediates to fill in the gaps 
between the seemingly discrete subgroups.

\section{Spectroscopic Subgroups}
\label{sec:subgroups}
For decades, SNe~Ia had been accepted for use as standard 
candles based on their apparent homogeneity, but in 1991 two very 
peculiar supernovae, SN~1991bg \citep{filippenko92, leibundgut93}
and SN~1991T \citep{waagen91, filippenko91T} challenged the 
validity of such use. 
According to work completed by \cite{li01} and recently updated 
\citep{li10} the peculiarity rate for SNe~Ia could be as high as 
30\%, with rates of 1991bg-likes up to 15\%. These observed 
peculiarities instigated the search for parameters 
which could be used to find a photometric and spectroscopic 
standard relation, in order to make SNe~Ia standardizable 
candles. 

The first relation was found by 
\cite{phillips93} who used a small sample of well-observed SNe~Ia 
to identify a very strong correlation between peak luminosity 
and initial decline rates, parameterized by $\Delta m_{15}$. 
The $\Delta m_{15}$ parameter measures the the decline in 
brightness of the {\emph B} band from maximum until $15$ days after 
maximum. The 
Phillips relation indicates brighter objects will have slower 
decline rates than dim objects which decline rapidly. An analogous 
spectroscopic relation was suggested by \cite{nugent95}.
Nugent's spectral relation measures the ratio of the strength 
of the Si~II features $\lambda$5972 and $\lambda$6355, denoted by 
$\mathcal{R}$(Si~II).  The standard relations, both photometric and 
spectroscopic, are due in large part to temperature differences 
caused by the amount of radioactive $^{56}$Ni synthesized during 
the explosion.

Single parameter modeling of SNe~Ia is indeed useful, but 
in cases of extremely peculiar SNe, such as SN 1991bg 
and SN 1991T, other means of accurate and efficient 
sub-classification are necessary. Several such classification 
schemes exist. One scheme proposed by \cite{benetti05a} is based on 
the evolution of Si~II $\lambda$6355 line velocity, while another, 
proposed by \cite{branch06} uses the pseudo-equivalent widths 
of the Si~II lines $\lambda$5972 and $\lambda$6355. 

By measuring 
the velocity of Si~II $\lambda$6355 10 days past {\emph B} max  
as well as the evolution of the velocity of this feature \citet{benetti05a}
found not only a spread in velocities between different SNe~Ia, but 
also a distribution in velocity gradients.  Plotting these gradients 
reveals three groups of SNe~Ia which \citet{benetti05a} labels as
faint (FAINT),  
high temporal velocity gradient (HVG), and low temporal velocity 
gradient (LVG) \citep{benetti05,benetti05a}.  The FAINT subgroup corresponds 
to subluminous 1991bg-like events with low expansion velocities 
and a large velocity gradient.  The HVG group contains normal 
SNe~Ia with high expansion velocities and a high velocity gradient. 
Finally, the LVG group contains both normal SNe~Ia as well as 
the brightest SNe~Ia.  This group has, on average, lower expansion 
velocities than the HVG group and a low velocity gradient.  

\begin{figure}
	\begin{center}
	\includegraphics[scale=0.5]{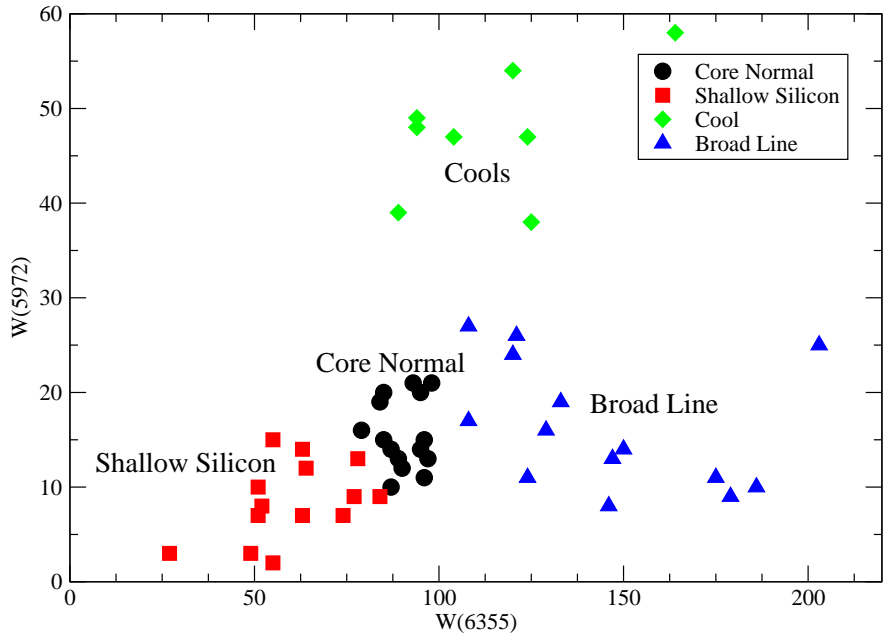}
	\caption{Pseudo-Equivalent Widths of $\lambda$5972 vs. 
		$\lambda$6355. Data comes from \cite{branch09} 
		as well as private communication
		with D. Branch.\label{ww4}}
	\end{center}
\end{figure}

In the scheme of \cite{branch06}, the pseudo-equivalent widths of 
Si~II lines $\lambda$5972 and $\lambda$6355 are measured and 
plotted (since emission lines are present in the spectra, these 
are not true equivalent widths but rather widths whose limits of 
integration have been chosen by eye). We followed the example of 
\cite{branch06} and plotted the width of the $\lambda$5972 feature vs. 
the width of the $\lambda$6355 feature for a sample of SNe~Ia. This plot
can be seen in Figure~\ref{ww4} and shows a cluster of core-normal (CN)
SNe~Ia with three discrete branches representing the peculiar 
spectroscopic subgroups of cool (CL), broad-line (BL), and 
shallow silicon (SS). The CLs correspond to Benetti's FAINT, BL 
tend to fall into the HVG group, and the LVG contains both CN and SS.  

In this method, each of the spectroscopic subgroups have their 
own differing properties and are named accordingly.  CNs 
are the most common and well studied SNe~Ia and are found in the 
region of highest density in the pseudo-equivalent width (WW) plot 
of Figure~\ref{ww4}. SN 1994D is an example of a CN SNe~Ia.  
BL SNe~Ia have Si~II lines whose absorptions are broader and deeper 
than CNs and thus form the lower right branch of the WW plot in 
Figure~\ref{ww4}. The SNe~Ia in the BL subgroup generally have 
the same ions that are present in CN, but with higher velocities. 
One such BL SNe~Ia is SN~2002bo \citep{branch06}. The 
SS SNe fall to the lower left of the plot in Figure~\ref{ww4} 
because of their smaller values for Si~II widths, but otherwise 
they have normal SNe~Ia spectra. (The CLs will be discussed in depth 
in \S\ref{sec:cools}.) Although the plot indicates there are 
separate spectroscopic subgroups the borders are not defined by a 
clear demarcation; instead some overlap exists. We call the SNe~Ia 
which lie between different subgroups ``intermediates".  

\section{Cool Type Ia Supernovae}
\label{sec:cools}
\begin{figure}
	\begin{center}
	\includegraphics[scale=0.5]{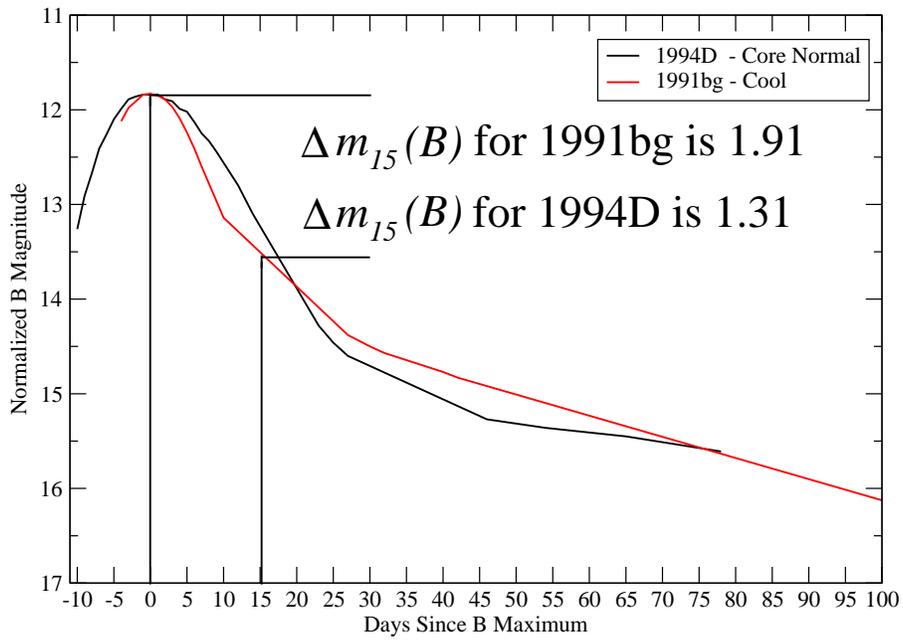}
	\caption{CN vs. CL Decline Rate Comparisons. $\Delta m_{15}$ 
		values originate from \cite{pastorello07}, light 
		curve data for 1991bg comes from \cite{turatto96}, 
		and light curve data for 1994D comes from \cite{patat96}.
		\label{normcoollc}}
	\end{center}
\end{figure}

\begin{figure}
	\begin{center}
	\includegraphics[scale=0.5]{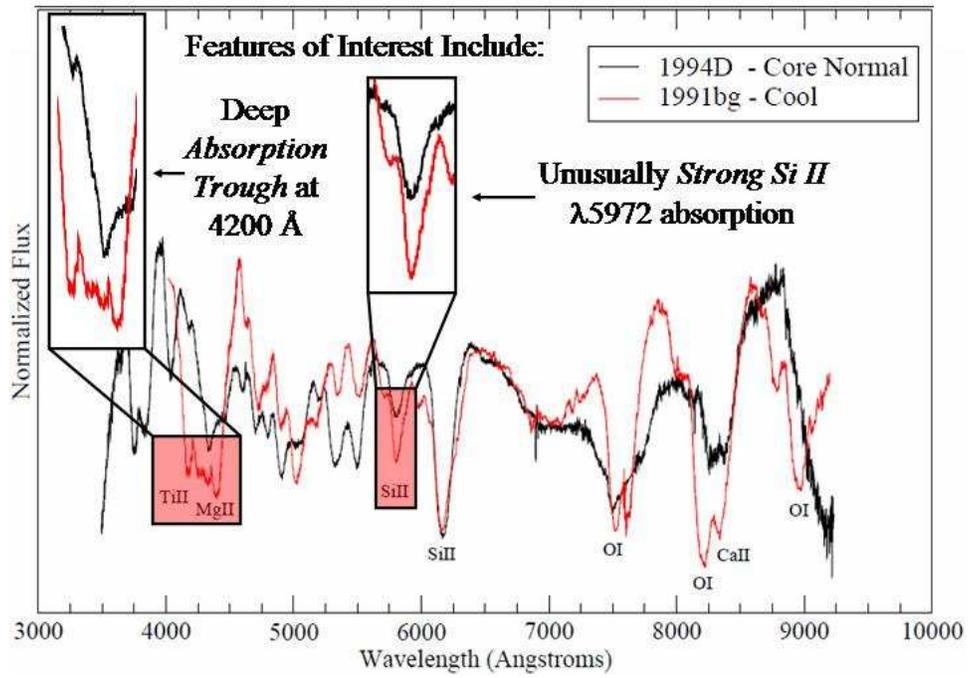}
	\caption{CN vs. CL Spectra Comparisons \label{normcoolsp}}
	\end{center}
\end{figure}

Spectroscopically peculiar SNe~Ia which fall into the cool subgroup 
share several distinct characteristics. Most notably, cools are 
subluminous. SN 1991bg had $V$ and $B$ maxima which were 
1.6 and 2.5 magnitudes, respectively, lower than ordinary 
SNe~Ia \citep{filippenko92}.  The CLs also have much faster 
declines from peak luminosity than do CNs.  Typically, CLs have
$\Delta m_{15}(B)\approx1.9$, while CNs have 
$\Delta m_{15}(B)\approx1.1$ \citep{taubenberger08}. Light curves of 
SN~1991bg and SN~1994D can be found in Figure~\ref{normcoollc} showing
appropriate $\Delta m_{15}(B)$ values.
It has been established that $^{56}$Ni production is fundamental to 
determining SNe~Ia peak luminosities \citep{arnett85, 
branch92}.  Analyses of SN~1991bg and other CLs have indicated 
$^{56}$Ni production of only about 1/6 the amount synthesized during a 
normal event \citep{mazzali97}. Spectroscopically, we see primarily 
that CLs have an unusually deep and wide absorption trough at around 
4200~\AA~and a very strong Si~II $\lambda5972$ absorption as shown in 
Figure~\ref{normcoolsp}.  The 4200~\AA~trough is well fit by 
Ti~II which can be explained by its low excitation temperature 
and the cool nature of these SNe~Ia \citep{filippenko92, mazzali97}. 
Finally, CLs also have a strong and narrow absorption at about 
5675~\AA~usually attributed to Na~I~D $\lambda5893$ 
\citep{filippenko92, garnavich04}.

\subsection{Progenitors}
\label{sec:prog}
To date, many theories and models have been proposed in attempts 
to describe the progenitors which lead to SNe~Ia, as well as 
to account for their diversity. Two common scenarios leading to 
SNe~Ia are generally accepted. One scenario is a single degenerate 
(SD) case proposed by \cite{whelan73} in which a white dwarf 
(WD) accretes mass until the Chandrasekhar Mass ($M_{Ch}$) is 
approached, igniting carbon in the core, undergoing thermonuclear 
instability, and ending with the complete disruption of the WD. 
A competing scenario proposed by \cite{iben84} as well as 
\cite{webbink84} involves two 
orbiting WDs, where loss of angular momentum due to the 
emission of gravitational waves leads to an inspiral and eventual 
merger. The merger of two WDs is referred to as double degenerate 
(DD). Extensive work has been done to investigate how either a SD 
scenario or a DD scenario can possibly explain both the 
homogeneity of SNe~Ia while at the same time account 
for the observed diversity. 

SNe~Ia appear to be homogeneous because the structure of the 
WD progenitor and the explosion is determined by nuclear physics 
\citep{hoeflich09}.  That is, the WD is supported by degenerate 
electron pressure, thermonuclear energy gives the energy production 
during explosion, and light curves are driven by the decay of 
synthesized radioactive $^{56}$Ni. According to \cite{hoeflich04}, 
the aspects most important to understanding the diversity are the 
conditions just prior to explosion and the type of burning mechanism. 
The basic types of burning can be distinguished as detonation and 
deflagration. Deflagration occurs with a burning front propagating 
due to heat transport across the front at a speed of $0.5-0.8\%$ the 
speed of sound \citep{timmes92}. On the other hand, in detonations 
compressional heat ignites carbon and oxygen in front of a shock 
which propagates supersonically \citep{hoeflich04}. 

SD models using delayed detonation---a process where 
deflagration occurs followed by a transition to detonation---show 
promise in reproducing observations \citep{khokhlov91}. 
Varying the amount of pre-expansion and burning during the 
deflagration phase as well as a few other secondary parameters 
can explain the diversity among SNe~Ia in a SD scenario 
\citep{hoeflich93, hoeflich09}. Namely, \cite{hoeflich09} decreases 
the density where transition from deflagration to detonation occurs 
and in so doing reduces the amount of $^{56}$Ni synthesized thus 
producing subluminous SNe~Ia. It should be noted that the trigger for 
transition from deflagration to detonation is still not fully understood.
When considering the DD scenario recent work 
 using the smoothed particle hydrodynamics 
code \texttt{GADGET3} shows promise in reproducing subluminous 
SNe~Ia akin to 1991bg-likes \citep{pakmor11,pakmor10}. They find that this 
type of merger reproduces the observed results 
and may be a path to the production of 1991bg-like SNe~Ia.

Another method used to investigate SNe~Ia progenitors is population
synthesis. Through this method, the observed delay time distribution 
(DTD) of SNe~Ia can be used to place limits upon progenitor models 
and formation scenarios. Taking an observational approach \cite{maoz10} 
derive the SN rates and DTD using supernova remnants in the 
Magellanic Clouds. \cite{maoz10} find that DTD is proportional 
to $t^{-1}$ in agreement with the DD DTD found from population synthesis 
of DD. \cite{maoz11} also confirmed this DTD result using star 
formation history. The general findings of population synthesis 
are that neither channel, SD or DD, can produce the observed number of 
SNe~Ia \citep{ruiter09, mennekens10}. \cite{mennekens10}, using the 
Brussels population synthesis code, find that the morphological shape 
of the observed DTD cannot be fit by the SD scenario alone.  Instead, 
a combination of both SD and DD scenarios will fit the morphological 
shape but produces approximately three times fewer SNe~Ia than is 
observed.  A similar study by \cite{ruiter09}, using the population 
synthesis code \texttt{StarTrack}, also find that the combination of 
SD and DD scenarios fits the observed shape of DTD, but accounts for 
ten times fewer SNe~Ia than observed. Population synthesis would 
then seem to indicate that neither scenario alone can produce the 
morphological shape of the DTD and that even the combination of both 
cannot produce the absolute number of observed SNe~Ia.

\section{Line Identifications in Cools}
\label{sec:id}
We begin our analysis of CL SNe~Ia by making line identifications.  
Identifying lines in SNe can be a difficult process and is 
further complicated in CLs due to the large number of ions with 
lower excitation temperatures and many metals which potentially are 
present as blends in the 4200~\AA~trough. Our sample contains 
spectra from SN 1991bg, SN 1997cn, SN 1999by, and SN 2005bl.
We find the nine ions---O~I, Na~I, Mg~II, Si~II, S~II, Ca~II, 
Ti~II, Cr~II, and Fe~II---of varying strengths and levels of 
stratification fit most of the observed spectral features well 
for all epochs considered with a tenth ion, Ca~I, present in 
the early spectra. 

\begin{figure}
	\begin{center}
	\includegraphics[scale=0.5]{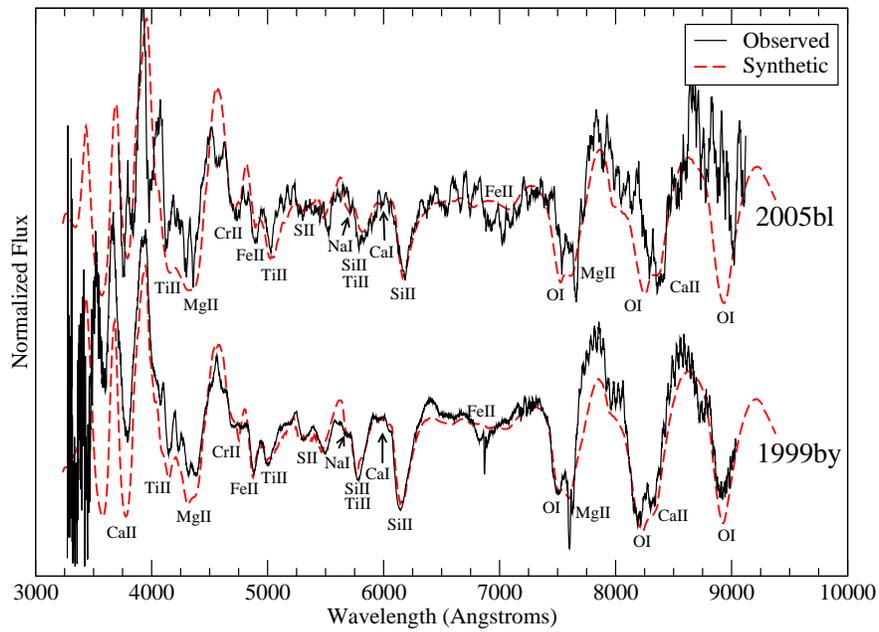}
	\caption{2005bl and 1999by Day 3 Pre {\emph B} Maximum. 
The telluric
                 oxygen feature at 7594~\AA\ is not marked.		 \label{day0-3comp}}
	\end{center}
\end{figure}

\begin{figure}
	\begin{center}
	\includegraphics[scale=0.5]{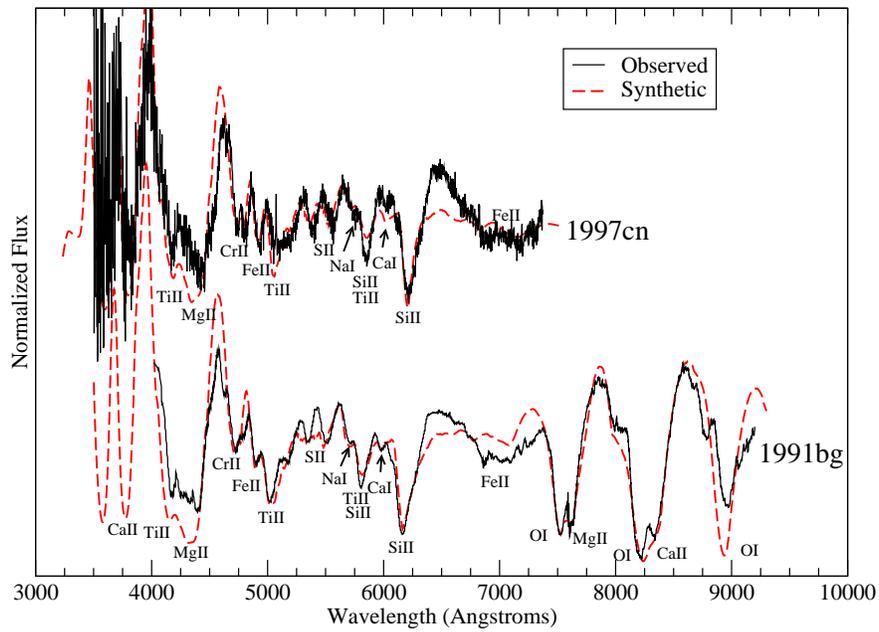}
	\caption{1997cn Day 3 Post {\emph B} Maximum and 
		 1991bg Day 2 Post {\emph B} Maximum. The telluric
                 oxygen feature at 7594~\AA\ is not marked.
		 \label{day003comp_a}}
	\end{center}
\end{figure}

\begin{figure}
	\begin{center}
	\includegraphics[scale=0.5]{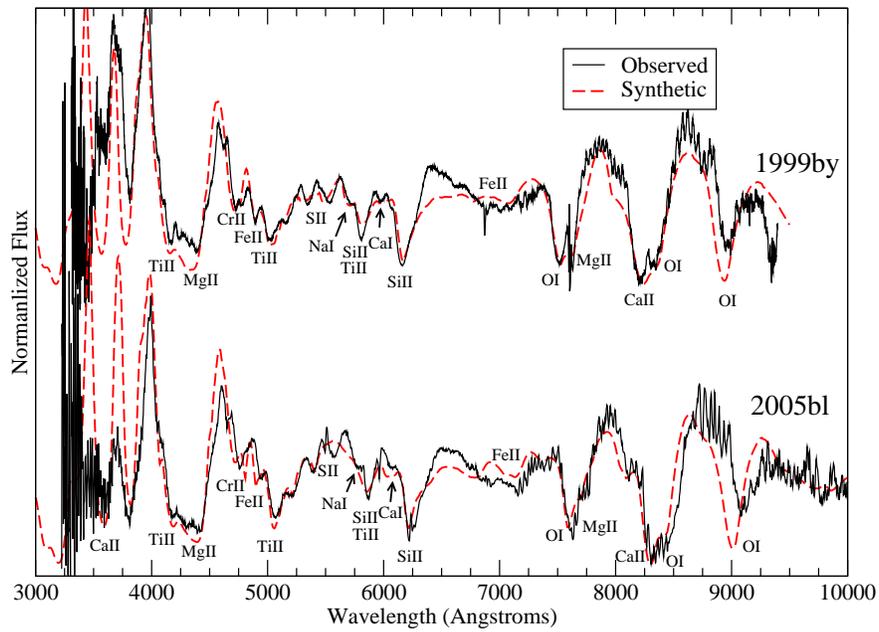}
	\caption{1999by Day 3 Post {\emph B} Maximum 
		 and 2005bl Day 4 Post {\emph B} Maximum.
The telluric
                 oxygen feature at 7594~\AA\ is not marked.
		 \label{day003comp_b}}
	\end{center}
\end{figure}

\begin{figure}
	\begin{center}
	\includegraphics[scale=0.5]{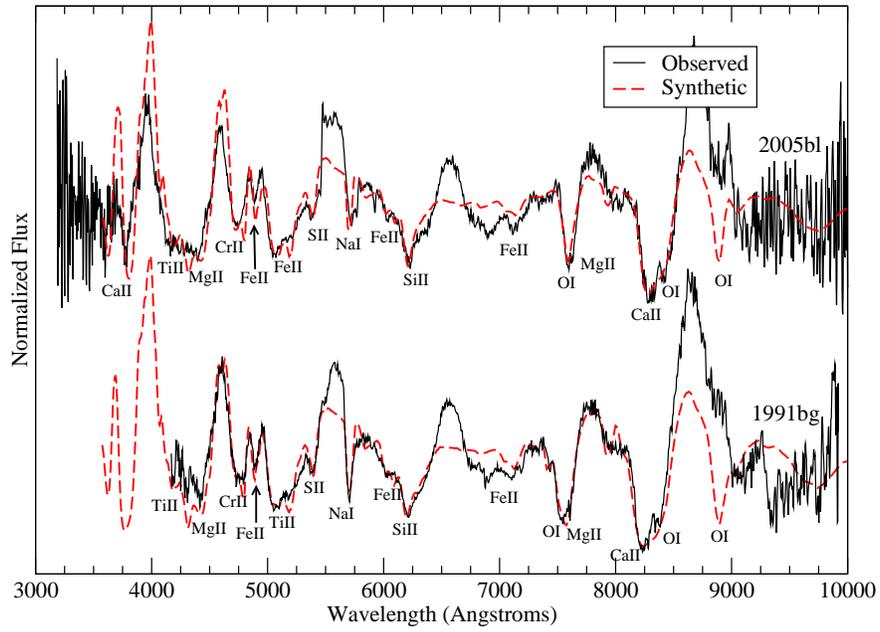}
	\caption{2005bl Day 19 Post {\emph B} Maximum, 
		 and 1991bg Day 18 Post {\emph B} Maximum. 
		 \label{day018comp}}
	\end{center}
\end{figure}

\begin{figure}
	\begin{center}
	\includegraphics[scale=0.5]{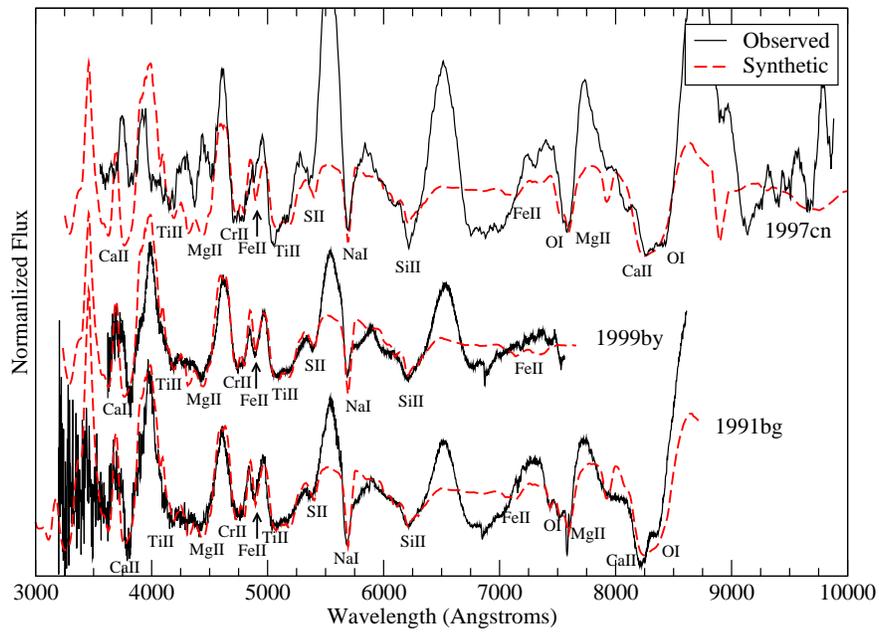}
	\caption{1997cn Day 28 Post {\emph B} Maximum,
		 1999by Day 29 Post {\emph B} Maximum, and
		 1991bg Day 29 Post {\emph B} Maximum.
		 \label{day028comp}}
	\end{center}
\end{figure}

\begin{figure}
	\begin{center}
	\includegraphics[scale=0.5]{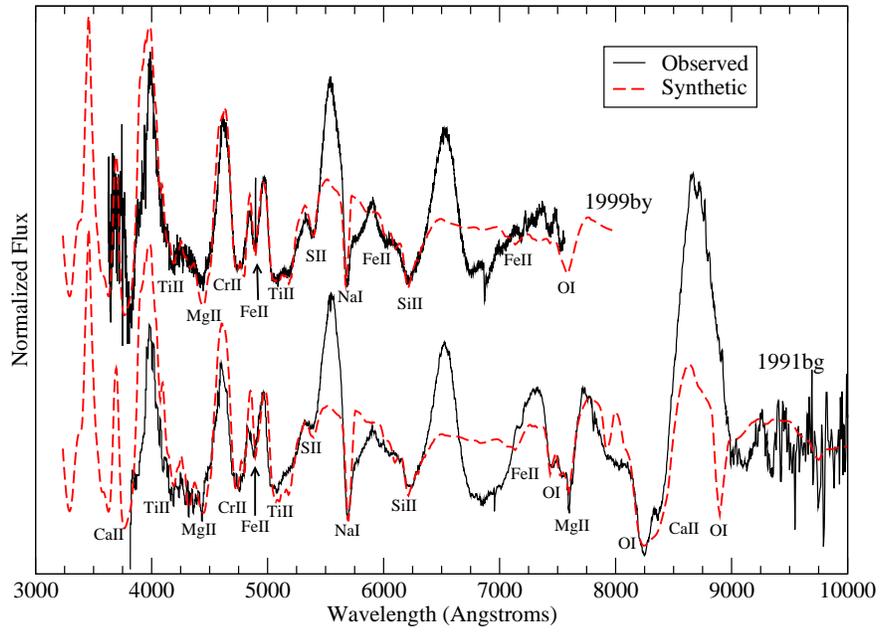}
	\caption{1999by Day 31 Post {\emph B} Maximum 
		 and 1991bg Day 32 Pre {\emph B} Maximum.
		 \label{day031comp}}
	\end{center}
\end{figure}

All with respect to {\emph B} maximum, we will make comparisons of 
five subsets of spectra from different epochs.  
In \S\ref{ssec:day-3} we compare SN~1999by and SN~2005bl at 
day $-3$.  The observed and synthetic spectra for this comparison are 
shown in Figure \ref{day0-3comp} and the fitting parameters are 
listed in Tables \ref{99byday0-3}--\ref{05blday0-3}. In \S\ref{ssec:day3} we analyze the spectra of SN~1991bg, SN~1997cn, 
SN~1999by, and SN~2005bl just past {\emph B} max. These spectra can be 
found in Figures \ref{day003comp_a}--\ref{day003comp_b} with fitting 
parameters listed in Tables \ref{91bgday002}--\ref{05blday004}. Synthetic spectra and line identifications for SN~1991bg at day 18 
and SN~2005bl at day 19 are summarized in \S\ref{ssec:day18} 
and displayed in Figure \ref{day018comp} with fitting parameters listed in 
Tables \ref{91bgday018}--\ref{05blday019}. In \S\ref{ssec:day28}, spectra from approximately one month 
post {\emph B} maximum for SN~1991bg, SN~1997cn, and SN~1999by 
are discussed.  The spectra at this epoch are plotted in Figures 
\ref{day028comp}--\ref{day031comp} with fitting parameters displayed 
in Tables \ref{97cnday028}--\ref{91bgday032}. 
A great deal of spectroscopic analysis has been performed on SN~1991bg
by \cite{filippenko92, leibundgut93, turatto96, mazzali97} and 
\cite{branch06, branch08, branch09}. 
\cite{turatto98} have analyzed SN~1997cn. \cite{garnavich04} and 
\cite{hoeflich02} considered SN~1999by. \cite{taubenberger08} 
and \cite{hachinger09} performed an analysis of SN~2005bl.

\subsection{Day 3 Pre {\emph B} Maximum Spectra}
\label{ssec:day-3}
The day $-3$ spectra are presented in Figure \ref{day0-3comp}.  The 
{\bf SYNOW} spectra for SN~1999by and SN~2005bl are overplotted with their 
respective observed spectra, and spectral absorptions are labeled. 
The synthetic spectra parameters are given in Tables 
\ref{99byday0-3}--\ref{05blday0-3}.

The broad absorption trough at 4200~\AA, characteristic of the CL 
subgroup, is well fit by a combination of Ti~II and Mg~II. This 
trough is continually fit by both Ti~II and Mg~II in all modeled 
spectra. \cite{garnavich04}, while modeling SN~1999by with {\bf SYNOW}, 
find the absorption at 5000~\AA~to be fit by Mg~I, whereas we 
attribute this feature to Ti~II and find no need of Mg~I.
Both Ti~II and Mg~I have an absorption at about 5000 \AA; however, SYNOW fits 
with Ti~II and without Mg~I give accurate fits for this absorption. Mg~I,
while possibly present, when included is not the main contributor to this 
feature and has no other absorptions or emissions which can clearly be 
attributed to it. Thus we consider the identification of Mg~I to not
be required.
 Ti~II also 
contributes to the 5780~\AA~feature.  The 7630~\AA~feature, 
blended into the O~I triplet, belongs to Mg~II.  For these reasons 
Ti~II and Mg~II can be considered definite. Other definite ions 
include Si~II, Ca~II, Fe~II, and O~I which all contribute to 
multiple absorptions.

The other four ions, Ca~I, Cr~II, S~II, and Na~I, only 
contribute to one absorption each and, although quite likely, are not definite. 
The ``W" feature at 5330~\AA, often attributed to S~II, may also 
be fit by features due to Sc~II; \cite{branch06} have indeed produced 
{\bf SYNOW} fits with this attribution.  Although synthetic fits with 
Sc~II are possible, S~II is more likely in the SNe Ia scenario, we
discuss this further below.
Na~I creates a shoulder in the 
5650~\AA~emission, and although very weak in the early spectra, 
becomes much more conspicuous later. Based on the evolution of multiple 
epochs, even though Na~I is weak early, it is highly likely to be 
present. Na~I is a common identification of this feature in CL SNe~Ia 
\citep{filippenko92, garnavich04, taubenberger08}; however, 
\cite{mazzali97} when using W7 cannot reproduce this feature 
without adding extra Na~I and modifying their abundance distributions, 
while \cite{leibundgut93} identify it as a blend of 
Si~II. Ca~I matches the absorption at 5980~\AA~in these 
spectra, but disappears sometime after day 4 post {\emph B} max and 
before day 18.  \cite{garnavich04} identifies 
two Ca~I absorptions as late as day 7 in SN~1999by and \cite{taubenberger08} 
make no Ca~I identifications for SN~2005bl. The remaining ion, 
Cr~II, is found at the 4700~\AA~absorption and further helps to 
block the emission at 4600~\AA, in agreement with \cite{taubenberger08}, 
but unidentified by \cite{garnavich04}.

Aside from \cite{garnavich04} identifying Mg~I in SN 1999by at 
5000~\AA~instead of Cr~II, our other line identifications are 
in good agreement. An analysis performed by \cite{hachinger09} 
on SN~2005bl identifies small amounts (less than $10\%$ of the 
mass fraction) of C~II in agreement with \cite{taubenberger08} 
who also find C~II. C~II is possible, however does not add any 
benefits to our synthetic fits and as noted by \cite{hachinger09} 
is a break from other CL SNe~Ia such as SN~1991bg and SN~1999by 
in which C~II has not been identified. In fact, C~II is more often 
associated with superluminous, rather than subluminous, SNe~Ia 
\citep{howell06, hicken07, parrent11}.

\subsection{Days 2, 3, and 4 Post {\emph B} Maximum Spectra}
\label{ssec:day3}
Figures \ref{day003comp_a}--\ref{day003comp_b} show the observed 
and synthetic spectra for SN~1991bg, SN~1997cn, SN~1999by and SN~2005bl 
at days 2, 3, 3, and 4, respectively.  The spectrum for SN~1997cn ends at 
7400~\AA, which is before the O~I feature and Mg~II absorption at 
7630~\AA; however these ions are still used in the {\bf SYNOW} fits 
as they contribute to the overall shape of the spectra. The fitting 
parameters used for these synthetic spectra are given in Tables 
\ref{91bgday002}--\ref{05blday004}.

As usual we consider Si~II, Ca~II, Fe~II, and O~I definitely 
present for their contribution to multiple spectral features. Also, 
Ti~II and Mg~II are definite ions except in the case of 1997cn where 
the Mg~II feature at 7630~\AA~is not available to be modeled. Due 
to the attribution of Mg~II to the absorption trough in SN~1997cn we still 
consider it highly likely.  Again, Cr~II continues to be present at 4700~\AA~and helps 
to reduce the emission just blueward of its absorption.  We consider 
the presence of Cr~II not only plausible but likely. Further, during 
these early spectra Ca~I fits the absorption at 5980~\AA.

In comparison with the analysis by \cite{turatto98} on SN~1997cn, we 
continue to see good agreement with the primary differences being 
Ca~I and Cr~II which they do not identify.  \cite{turatto98} also 
identifies Ni~II at 4067~\AA, which is possible, but we find no 
benefit in the inclusion of Ni~II in our fits.  

Close examination of the fitting parameters given for these SNe in 
Tables \ref{91bgday002}--\ref{05blday004} will show that both 
1997cn and 2005bl have photospheric velocities ($v_{phot}$) of 
$7,600~\mathrm{km~s}^{-1}$ which is 
significantly lower than 1991bg at $11,000~\mathrm{km~s}^{-1}$ and 1999by at 
$10,000~\mathrm{km~s}^{-1}$.  Looking forward to days 18 and 19 we see that 
2005bl has a lower gradient in $v_{phot}$ than does 1991bg and 
differs by only $100~\mathrm{km~s}^{-1}$. By days 28, and 29 the differences 
in $v_{phot}$ between the SNe are insignificant.  

\begin{figure}
	\begin{center}
	\includegraphics[scale=0.5]{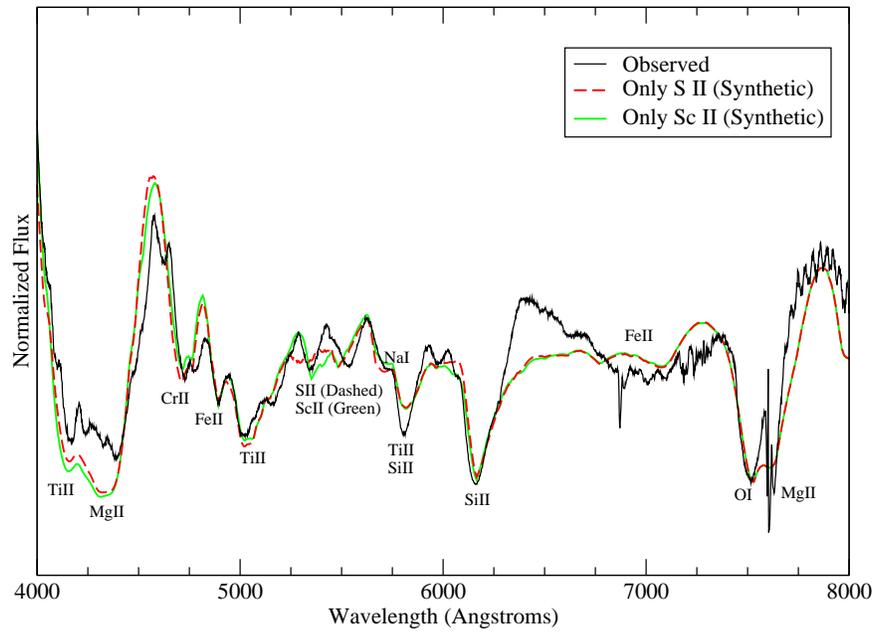}
	\caption{1999by Day 3 Post {\emph B} Maximum with the feature
          usually attributed to S~II fit with both S~II and Sc~II. 
The telluric
                 oxygen feature at 7594~\AA\ is not
                 marked.		 \label{fig:day3_sc2_s2_comp}} 
	\end{center}
\end{figure}

Figure~\ref{fig:day3_sc2_s2_comp} compares the fit of SN~1999by at day
+3, using S~II and Sc~II to fit the ``W'' feature usually attributed
to S~II. It is hard to definitively decide between the two fits and
without the strong S~II $\lambda 6355$ line, the identification of
SN~1999by as a SNe Ia becomes weaker. This effect is important but
needs to be followed up in future work with more detailed spectral
modeling.

\subsection{Days 18 and 19 Post {\emph B} Maximum Spectra}
\label{ssec:day18}
All the same ions as discussed in 
\S\S\ref{ssec:day-3}--\ref{ssec:day3}, except Ca~I, are still present 
by days 18 and 19, but the spectra have several noticeable 
differences from the early time spectra. (Synthetic and observed spectra 
for SN~1991bg and SN~2005bl can be seen in Figure \ref{day018comp}.) 
Primarily the differences include the ions having 
deeper absorptions, higher emissions, and lower $v_{phot}$. 
The most obvious difference is the Na~I absorption at 5700~\AA, which is 
no longer just a shoulder in an emission but a rather strong and 
sharp absorption. We still find Si~II, Ca~II, Fe~II, Ti~II, Mg~II, 
and O~I definite whereas Na~I is now much more likely.  Cr~II 
and S~II are plausible for the same reasons as before and Ca~I is 
no longer used in the synthetic spectra.

\subsection{Days 28, 29, 31, and 32 Post {\emph B} Maximum Spectra}
\label{ssec:day28}
Figures \ref{day028comp}--\ref{day031comp} have the synthetic and 
observed spectra for SN~1991bg, SN~1997cn, and SN~1999by where the absorptions 
are well fit; however, in the red the synthetic emissions are weak. This 
lack of strong emissions can be explained by the assumptions of {\bf SYNOW} 
itself. {\bf SYNOW} assumes a perfect blackbody emitter and absorber 
with a sharp photosphere surrounded by a homologous and spherically 
expanding ion cloud where lines form by resonance scattering treated 
in the Sobolev approximation. These assumptions do not allow for 
net emission profiles. Once ions begin to enter net emission 
{\bf SYNOW} can no longer reproduce the strong emission peaks well. 
To overcome this issue it is possible to increase greatly the optical
depth and rescale, but better fits are often obtained simply by 
leaving these emissions alone and focusing on the absorptions.

The same nine ions are present in these spectra as the previous, but 
O~I has been strongly stratified to fit the three separate absorptions 
at about 7530~\AA. Furthermore, a deep and wide absorption around 
6700~\AA~to 7000~\AA~has appeared. Despite our best efforts we were 
unable to fit this trough in a reasonable way.  We tried 
a variety of ions including C~I and O~II, which could fill some of the 
trough but not without heavy restriction in $v_{min}$ and $v_{max}$. 
None of the attempts proved successful and we do not find any of these 
ions likely.

\section{Intermediate Type Ia Supernovae}
\label{sec:ints}
\begin{figure}
	\begin{center}
	\includegraphics[scale=0.5]{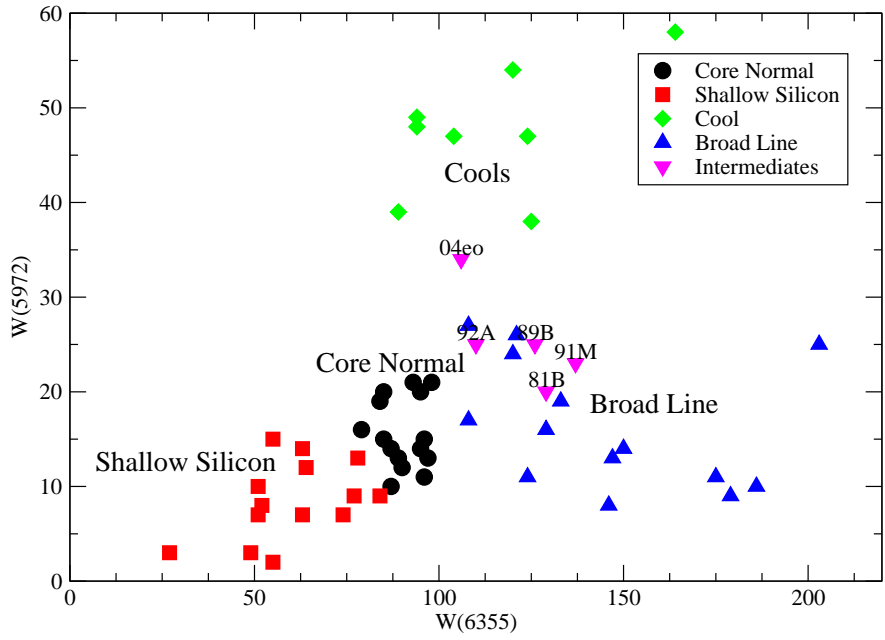}
	\caption{Pseudo-Equivalent Widths of $\lambda$5972 vs. 
		$\lambda$6355 including intermediates. Data comes from 
		\cite{branch09} as well as private communication
		with D. Branch. \label{ww5}}
	\end{center}
\end{figure}

Figure \ref{ww5} is another WW plot very similar to Figure \ref{ww4}, 
but in this WW plot, intermediates are included.  
Intermediates are SNe~Ia that do not necessarily fall into any 
particular subgroup; instead they fall in between, sharing 
properties of the different subgroups. What we find when including 
the intermediates is that the subgroups begin to blend, indicating 
that rather than being discrete, SNe~Ia subgroups, as delineated 
by Branch, are a continuous distribution based on multiple parameters.  
\cite{branch06, branch09} did extensive work on spectroscopic 
subclassification of SNe~Ia and came to a similar conclusion.

Recent work of \cite{maeda10} indicates that some of the spectral diversity 
of SNe~Ia is a result of the viewing angle on an asymmetrically 
exploding WD progenitor. \cite{maeda10} used the Benetti classification 
scheme of FAINT, HVG, and LVG.  Using the emission lines 
Fe~II~$\lambda 7155$ and Ni~II~$\lambda 7378$ lines, which occur 
during the nebular phase, they measured the Doppler shift in the 
ejecta. Blue-shifted lines will be on the near side and red-shifted will 
be on the far side of an exploding SN. They found a diversity in the 
velocities of the blue-shifted and red-shifted lines implying that the 
initial SN ignition occurred off-center. Measurement of red-shift and 
blue-shift in HVG and LVG SNe~Ia show that HVGs are preferentially 
red-shifted and LVGs blue-shifted. \cite{maeda10} conclude that LVGs 
are viewed in the direction of the initial spark and HVGs are viewed 
opposite the spark. Statistical treatment of this result indicates a 
high probability that the differences of the two groups is entirely an 
effect of viewing angle. The work of \cite{maeda10} explained why the 
differences between the HVG and LVG groups occurred, but did little to 
address the FAINT group which they acknowledge may arise from a completely 
different explosion mechanism then HVGs and LVGs.

\section{Discussion}
\label{sec:disc}
Due to the high level of importance placed on SNe~Ia, a complete 
understanding of their physical nature is paramount. One method to probe 
this nature is examination of their differences. Thus, we analyzed the 
spectroscopically CL subgroup of SNe~Ia. In summary, SNe in this subgroup 
are underluminous, rapid decliners, and poor $^{56}$Ni producers with deep 
and wide absorption troughs around 4200~\AA~due to Ti~II and unusually strong 
Si~II $\lambda5972$ absorptions. The path leading to subluminous SNe~Ia is 
presently uncertain. SD scenarios in which a delayed-detonation, where a subsonic 
deflagration front transitions into a detonation, are strong theoretically 
and account well for observations \citep{hoeflich09}. There also exist 
competing DD scenarios such as that presented by \cite{pakmor11} which 
indicate SNe~Ia potentially originate from the merger of two CO WDs. 
Each scenario has its own strengths and weaknesses and the question 
of progenitors remains open. Furthermore, population synthesis studies 
tell us that neither scenario alone can account for the total number of 
observed SNe~Ia.

We have modeled the spectra of several CL SNe~Ia with the fast and 
highly parameterized synthetic spectrum code {\bf SYNOW} and 
compared our findings with those of similar studies.  We find good 
agreement with other studies.  Our work indicates many ions are 
potentially present in CLs, and line identifications are complicated 
for this reason.  We show that Si~II, Ca~II, Fe~II, Ti~II, Mg~II and O~I 
are the most definite ions present in the spectra of cools.  It is also 
very likely that Na~I is present.  Three other ions, Ca~I, Cr~II, and 
S~II, are considered as possibilities. S~II is quite likely, but Ca~I 
and Cr~II only occur in one location in our synthetic spectra and they 
have not been definitively explained in other studies. Therefore we consider 
these ions possible but not definite. 

Examining the velocity evolution 
of Si~II~$\lambda6355$ shows the primary difference among this subgroup.  
From our models we see that SN~2005bl and SN~1999by have lower 
Si~II~$\lambda6355$ velocities than SN~1991bg and SN~1997cn in early 
time spectra but have lower velocity gradients so that by about day 18 
post {\emph B} maximum all four CLs have approximately the same velocity.  
Si~II~$\lambda6355$ velocities are plotted in Figure~\ref{si2vels} for 
several CLs and CNs. Comparing our velocities with those of other studies 
show insignificant differences. Overall, the appearance of the velocity 
distribution for all SNe~Ia is quite continuous.

\begin{figure}
	\begin{center}
	\includegraphics[scale=1.5]{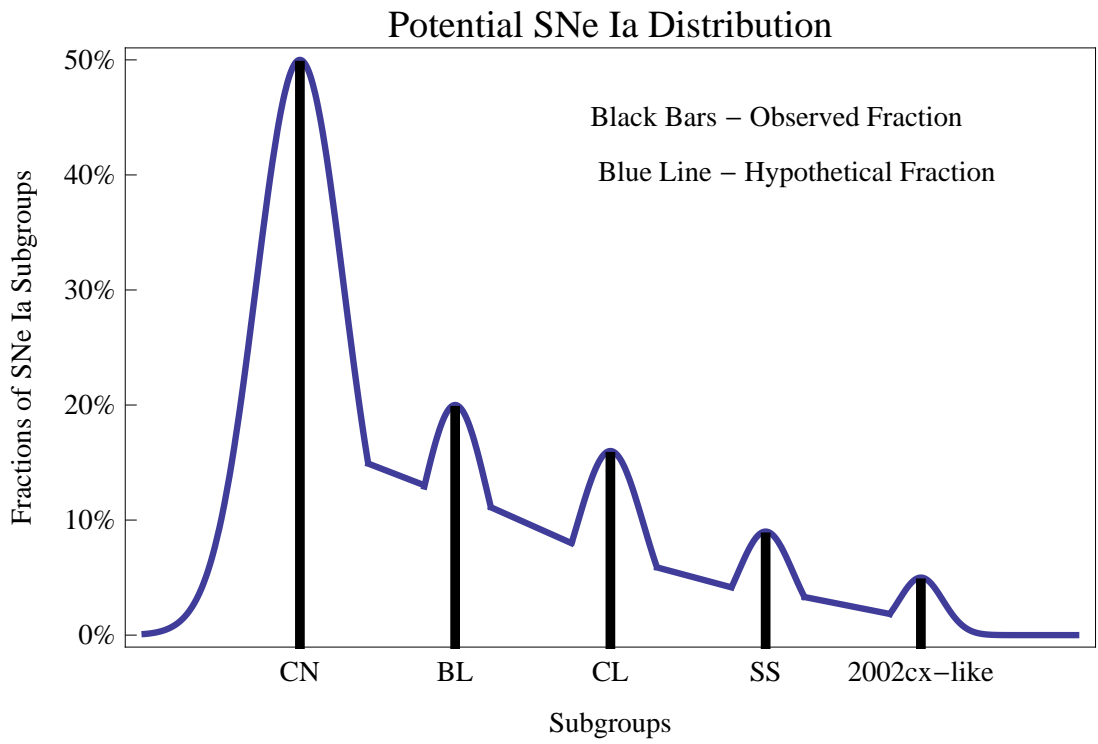}
	\caption{Cartoon SNe~Ia distribution.  The observed 
		subgroup fraction data comes from \cite{li10}. 
		\label{histogram} }
	\end{center}
\end{figure}

Our examination of this subgroup indicates that 1991bg-likes 
are fairly homogeneous amongst themselves (albeit somewhat due 
to construction) and not only share the same ions, but fitting 
parameters and properties as well. Upon consideration of intermediates 
the distinctions among the subgroups begin to blur even more and the 
distribution of SNe~Ia seems increasingly continuous. With a large 
and well-observed sample this distribution may take a form similar to the 
hypothetical distribution displayed in Figure \ref{histogram}.

A continuous distribution of SNe~Ia could have regions of high density 
and low density in the WW plots, but the difference in density should 
be fairly small. These dense regions would contain the subgroups. 
The highest density region would contain the most common SNe~Ia, 
the CNs. The lowest density regions would contain SNe~Ia which 
are steps away from one subgroup toward another; intermediates. 
The reason for different regions of high density and low 
density, i.e. the subgroups, could indicate sets of secondary 
parameters which are slightly more stable and thus more likely 
to be reproduced for WDs transitioning into SNe~Ia, rather than 
fundamentally different processes.    

Present work on the diversity of SNe~Ia is incomplete but is 
leading to the conclusion that rather than being made of discrete 
subgroups, SNe~Ia are a continuous distribution of the same phenomenon. 
However, for more definitive conclusions it is necessary to have a 
much larger and well-observed sample coupled with more 
robust explosion models.

\section{Acknowledgments}
\label{sec:ack}
We thank David Branch for many helpful discussions, Saurabh 
Jha for providing his data, and Stefano Benetti for providing 
his data. We thank the anonymous referee for improving our
presentation and bringing the S~II versus Sc~II identification
question into clearer focus.

This work was supported in part by NSF grant AST-0707704,  US DOE 
Grant DE-FG02-07ER41517, and NASA Grant HST-GO-12298.05-A.
Support for Program number HST-GO-12298.05-A was provided by 
NASA through a grant from the Space Telescope Science Institute, 
which is operated by the Association of Universities for Research 
in Astronomy, Incorporated, under NASA contract NAS5-26555.

\bibliographystyle{apj}
\bibliography{doull_1991bg}

\clearpage{}

\clearpage{}

\clearpage

\clearpage

\clearpage

\begin{deluxetable}{lcccccc}
	\tabletypesize{\scriptsize}
	\tablecaption{Fitting Parameters for 1999by Day 3 Pre {\emph B} Maximum.
			\label{99byday0-3}}
	\tablewidth{0pt}
	\tablehead{
		\colhead{Element}         & \colhead{$\tau$}
		& \colhead{$T_{exc}$ (K)} & \colhead{$v_{phot}$} 
		& \colhead{$v_{e}$}    & \colhead{$v_{min}$}
		& \colhead{$v_{max}$} }
	
	\startdata
	H$V_{e}$~O I& 4.5  & 7000 & 11 & 1.1 & 0    & 40    \\
	O I   & -    & -    & -  & -   & -    & -     \\
	Na I  & 1.1  & 7000 & 11 & 0.5 & 12.5 & 15    \\
	Mg II & 20   & 7000 & 11 & 1.0 & 0    & 40    \\
	Si II & 550  & 7000 & 11 & 0.5 & 0    & 13.5  \\
	S II  & 12   & 7000 & 11 & 0.2 & 0    & 40    \\
	Ca I  & 8.0  & 7000 & 10 & 1.0 & 0    & 40    \\
	Ca II & 800  & 7000 & 11 & 1.2 & 0    & 40    \\
	Ti II & 2    & 7000 & 11 & 2.0 & 0    & 40    \\
	Cr II & 600  & 7000 & 11 & 1.0 & 0    & 40    \\
	Fe II & 3.0  & 5000 & 11 & 1.0 & 0    & 40    \\
	\enddata
\tablenotetext{*}{All velocities are given in $1000~\mathrm{km~s}^{-1}$}
\end{deluxetable}

\begin{deluxetable}{lcccccc}
	\tabletypesize{\scriptsize}
	\tablecaption{Fitting Parameters for 2005bl Day 3 Pre {\emph B} Maximum.
			\label{05blday0-3}}
	\tablewidth{0pt}
	\tablehead{
		\colhead{Element}         & \colhead{$\tau$}
		& \colhead{$T_{exc}$ (K)} & \colhead{$v_{phot}$} 
		& \colhead{$v_{e}$}    & \colhead{$v_{min}$}
		& \colhead{$v_{max}$} }
	
	\startdata
	H$V_{e}$~O I& 4.5 & 7000 & 10 & 1.4 & 0    & 40   \\
	O I   &  -  & -    & -  & -   & -    & -    \\
	Na I  & 0.9 & 7000 & 10 & 0.5 & 12.5 & 14.5 \\
	Mg II & 40  & 7000 & 10 & 1.0 & 0    & 40   \\
	Si II & 80  & 7000 & 10 & 0.5 & 0    & 40   \\
	S II  & 8.0 & 7000 & 10 & 0.2 & 0    & 40   \\
	Ca I  & 8.0 & 7000 & 10 & 1.0 & 0    & 40   \\
	Ca II & 1000& 7000 & 10 & 1.0 & 0    & 40   \\
	Ti II & 6.0 & 7000 & 10 & 2.0 & 0    & 40   \\
	Cr II & 1200& 7000 & 10 & 1.0 & 0    & 40   \\
	Fe II & 3.0 & 7000 & 10 & 1.0 & 0    & 40   \\
	\enddata
\tablenotetext{*}{All velocities are given in $1000~\mathrm{km~s}^{-1}$}
\end{deluxetable}
\clearpage

\clearpage

\begin{deluxetable}{lcccccc}
	\tabletypesize{\scriptsize}
	\tablecaption{Fitting Parameters for 1991bg Day 2 Post {\emph B} Maximum.
			\label{91bgday002}}
	\tablewidth{0pt}
	\tablehead{
		\colhead{Element}         & \colhead{$\tau$}
		& \colhead{$T_{exc}$ (K)} & \colhead{$v_{phot}$} 
		& \colhead{$v_{e}$}    & \colhead{$v_{min}$}
		& \colhead{$v_{max}$} }
	
	\startdata
	H$V_{e}$~O I& 5.0  & 7000 & 10 & 1.4 & 0    & 40   \\
	O I   &   -  & -    & -  & -   & -    & -    \\
	Na I  & 0.5  & 7000 & 10 & 1.0 & 12.5 & 13.5 \\
	Mg II & 40.0 & 7000 & 10 & 1.0 & 0    & 40   \\
	Si II & 100  & 7000 & 10 & 0.5 & 0    & 40   \\
	S II  & 5.0  & 7000 & 10 & 0.2 & 0    & 40   \\
	Ca I  & 15.0 & 7000 & 10 & 1.0 & 0    & 40    \\
	Ca II & 1100 & 7000 & 10 & 1.3 & 0    & 40   \\
	Ti II & 6.0  & 7000 & 10 & 2.0 & 0    & 40   \\
	Cr II & 1500 & 7000 & 10 & 1.0 & 0    & 40   \\
	Fe II & 4.0  & 7000 & 10 & 1.0 & 0    & 40   \\
	\enddata
\tablenotetext{*}{All velocities are given in $1000~\mathrm{km~s}^{-1}$}
\end{deluxetable}

\begin{deluxetable}{lcccccc}
	\tabletypesize{\scriptsize}
	\tablecaption{Fitting Parameters for 1997cn Day 3 Post {\emph B} Maximum.
			\label{97cnday003}}
	\tablewidth{0pt}
	\tablehead{
		\colhead{Element}         & \colhead{$\tau$}
		& \colhead{$T_{exc}$ (K)} & \colhead{$v_{phot}$} 
		& \colhead{$v_{e}$}    & \colhead{$v_{min}$}
		& \colhead{$v_{max}$} }
	
	\startdata
	H$V_{e}$~O I& 5.0  & 7000 & 7.6 & 1.4 & 0    & 40   \\
	O I   &   -  & -    & -   & -   & -    & -    \\
	Na I  & 0.1  & 7000 & 7.6 & 1.0 & 10.5 & 17.5 \\
	Mg II & 40.0 & 7000 & 7.6 & 1.0 & 0    & 40   \\
	Si II & 30.0 & 7000 & 7.6 & 0.6 & 0    & 40   \\
	S II  & 5.0  & 7000 & 7.6 & 0.2 & 0    & 40   \\
	Ca I  &  8.0 & 7000 & 10 & 1.0 & 0    & 40    \\
	Ca II & 1100 & 7000 & 7.6 & 1.3 & 0    & 40   \\
	Ti II & 4.0  & 7000 & 7.6 & 2.0 & 0    & 40   \\
	Cr II & 1000 & 7000 & 7.6 & 1.0 & 0    & 40   \\
	Fe II & 6.0  & 7000 & 7.6 & 1.0 & 0    & 40   \\
	\enddata
\tablenotetext{*}{All velocities are given in $1000~\mathrm{km~s}^{-1}$}
\end{deluxetable}
\clearpage

\clearpage

\clearpage

\begin{deluxetable}{lcccccc}
	\tabletypesize{\scriptsize}
	\tablecaption{Fitting Parameters for 1999by Day 3 Post {\emph B} Maximum.
			\label{99byday003}}
	\tablewidth{0pt}
	\tablehead{
		\colhead{Element}         & \colhead{$\tau$}
		& \colhead{$T_{exc}$ (K)} & \colhead{$v_{phot}$} 
		& \colhead{$v_{e}$}    & \colhead{$v_{min}$}
		& \colhead{$v_{max}$} }
	
	\startdata
	H$V_{e}$~O I& 5.0  & 7000 & 10 & 1.4 & 0    & 40   \\
	O I   &   -  & -    & -  & -   & -    & -    \\
	Na I  & 0.5  & 7000 & 10 & 1.0 & 12.5 & 13.5 \\
	Mg II & 40.0 & 7000 & 10 & 1.0 & 0    & 40   \\
	Si II & 100  & 7000 & 10 & 0.5 & 0    & 40   \\
	S II  & 5.0  & 7000 & 10 & 0.2 & 0    & 40   \\
	Ca I  & 12.0 & 7000 & 10 & 1.0 & 0    & 40    \\
	Ca II & 1100 & 7000 & 10 & 1.3 & 0    & 40   \\
	Ti II & 6.0  & 7000 & 10 & 2.0 & 0    & 40   \\
	Cr II & 1500 & 7000 & 10 & 1.0 & 0    & 40   \\
	Fe II & 4.0  & 7000 & 10 & 1.0 & 0    & 40   \\
	\enddata
\tablenotetext{*}{All velocities are given in $1000~\mathrm{km~s}^{-1}$}
\end{deluxetable}

\begin{deluxetable}{lcccccc}
	\tabletypesize{\scriptsize}
	\tablecaption{Fitting Parameters for 2005bl Day 4 Post {\emph B} Maximum.
			\label{05blday004}}
	\tablewidth{0pt}
	\tablehead{
		\colhead{Element}         & \colhead{$\tau$}
		& \colhead{$T_{exc}$ (K)} & \colhead{$v_{phot}$} 
		& \colhead{$v_{e}$}    & \colhead{$v_{min}$}
		& \colhead{$v_{max}$} }
	
	\startdata
	H$V_{e}$~O I&   -  & -    & -   & -   & -    & -    \\
	O I   & 4.5  & 7000 & 7.6 & 1.0 & 0    & 40   \\
	Na I  & 0.1  & 7000 & 7.6 & 1.0 & 10   & 12   \\
	Mg II & 10.0 & 7000 & 7.6 & 1.0 & 0    & 40   \\
	Si II & 10.0 & 7000 & 7.6 & 0.6 & 0    & 40   \\
	S II  & 1.0  & 7000 & 7.6 & 0.2 & 0    & 40   \\
	Ca I  &  8.0 & 7000 & 10 & 1.0 & 0    & 40    \\
	Ca II & 800  & 7000 & 7.6 & 1.3 & 0    & 40   \\
	Ti II & 7.0  & 7000 & 7.6 & 2.0 & 0    & 40   \\
	Cr II & 600  & 7000 & 7.6 & 1.0 & 0    & 40   \\
	Fe II & 6.0  & 7000 & 7.6 & 1.0 & 0    & 40   \\
	\enddata
\tablenotetext{*}{All velocities are given in $1000~\mathrm{km~s}^{-1}$}
\end{deluxetable}
\clearpage

\clearpage

\begin{deluxetable}{lcccccc}
	\tabletypesize{\scriptsize}
	\tablecaption{Fitting Parameters for 1991bg Day 18 Post {\emph B} Maximum.
			\label{91bgday018}}
	\tablewidth{0pt}
	\tablehead{
		\colhead{Element}         & \colhead{$\tau$}
		& \colhead{$T_{exc}$ (K)} & \colhead{$v_{phot}$} 
		& \colhead{$v_{e}$}    & \colhead{$v_{min}$}
		& \colhead{$v_{max}$} }
	
	\startdata
	H$V_{e}$~O I& 2.0  & 7000 & 8 & 1.6 & 0    & 40   \\
	O I   & 0.7  & 7000 & 8 & 1.0 & 15.2 & 16   \\
	Na I  & 9.0  & 7000 & 8 & 0.5 & 11   & 12   \\
	Mg II & 3.5  & 7000 & 8 & 1.0 & 12   & 40   \\
	Si II & 50   & 7000 & 8 & 0.5 & 0    & 40   \\
	S II  & 0.8  & 7000 & 8 & 1.0 & 14.5 & 40   \\
	Ca II & 4000 & 7000 & 8 & 1.3 & 0    & 40   \\
	Ti II & 12   & 7000 & 8 & 1.0 & 0    & 40   \\
	Cr II & 2500 & 7000 & 8 & 1.0 & 0    & 40   \\
	Fe II & 70   & 7000 & 8 & 0.7 & 0    & 40   \\
	\enddata
\tablenotetext{*}{All velocities are given in $1000~\mathrm{km~s}^{-1}$}
\end{deluxetable}

\begin{deluxetable}{lcccccc}
	\tabletypesize{\scriptsize}
	\tablecaption{Fitting Parameters for 2005bl Day 19 Post {\emph B} Maximum.
			\label{05blday019}}
	\tablewidth{0pt}
	\tablehead{
		\colhead{Element}         & \colhead{$\tau$}
		& \colhead{$T_{exc}$ (K)} & \colhead{$v_{phot}$} 
		& \colhead{$v_{e}$}    & \colhead{$v_{min}$}
		& \colhead{$v_{max}$} }
	
	\startdata
	H$V_{e}$~O I& 1.0  & 7000 & 7.5 & 1.6 & 0    & 40   \\
	O I   &   -  & -    & -   & -   & -    & -    \\
	Na I  & 3.0  & 7000 & 7.5 & 0.5 & 11   & 12   \\
	Mg II & 2.0  & 7000 & 7.5 & 1.0 & 12   & 40   \\
	Si II & 40   & 7000 & 7.5 & 0.5 & 0    & 40   \\
	S II  & 0.7  & 7000 & 7.5 & 1.0 & 14.5 & 40   \\
	Ca II & 4000 & 7000 & 7.5 & 1.0 & 0    & 40   \\
	Ti II & 17   & 7000 & 7.5 & 1.0 & 0    & 40   \\
	Cr II & 2800 & 7000 & 7.5 & 1.0 & 0    & 40   \\
	Fe II & 30   & 7000 & 7.5 & 0.8 & 0    & 40   \\
	\enddata
\tablenotetext{*}{All velocities are given in $1000~\mathrm{km~s}^{-1}$}
\end{deluxetable}
\clearpage

\clearpage

\begin{deluxetable}{lcccccc}
	\tabletypesize{\scriptsize}
	\tablecaption{Fitting Parameters for 1997cn Day 28 Post {\emph B} Maximum.
			\label{97cnday028}}
	\tablewidth{0pt}
	\tablehead{
		\colhead{Element}         & \colhead{$\tau$}
		& \colhead{$T_{exc}$ (K)} & \colhead{$v_{phot}$} 
		& \colhead{$v_{e}$}    & \colhead{$v_{min}$}
		& \colhead{$v_{max}$} }
	
	\startdata
	H$V_{e}$~O I& 0.6  & 7000 & 7.4 & 1.6 & 0    & 40   \\
	O I   & 0.4  & 7000 & 7.4 & 1.0 & 14   & 16   \\
	Na I  & 6.0  & 7000 & 7.4 & 0.7 & 11   & 40   \\
	Mg II & 7.0  & 7000 & 7.4 & 1.0 & 12   & 13   \\
	Si II & 10   & 7000 & 7.4 & 0.5 & 0    & 40   \\
	S II  & 0.4  & 7000 & 7.4 & 1.0 & 14   & 40   \\
	Ca II & 4000 & 7000 & 7.4 & 1.3 & 0    & 40   \\
	Ti II & 8.0  & 7000 & 7.4 & 1.0 & 0    & 40   \\
	Cr II & 3000 & 7000 & 7.4 & 1.0 & 0    & 40   \\
	Fe II & 50   & 7000 & 7.4 & 0.6 & 0    & 40   \\
	\enddata
\tablenotetext{*}{All velocities are given in $1000~\mathrm{km~s}^{-1}$}
\end{deluxetable}

\begin{deluxetable}{lcccccc}
	\tabletypesize{\scriptsize}
	\tablecaption{Fitting Parameters for 1991bg Day 29 Post {\emph B} Maximum.
			\label{91bgday029}}
	\tablewidth{0pt}
	\tablehead{
		\colhead{Element}         & \colhead{$\tau$}
		& \colhead{$T_{exc}$ (K)} & \colhead{$v_{phot}$} 
		& \colhead{$v_{e}$}    & \colhead{$v_{min}$}
		& \colhead{$v_{max}$} }
	
	\startdata
	H$V_{e}$~O I& 0.6  & 7000 & 7.6 & 1.6 & 0    & 40   \\
	O I   & 0.4  & 7000 & 7.6 & 1.0 & 14   & 16   \\
	Na I  & 6.0  & 7000 & 7.6 & 0.7 & 11   & 40   \\
	Mg II & 7.0  & 7000 & 7.6 & 1.0 & 12   & 13   \\
	Si II & 10   & 7000 & 7.6 & 0.5 & 0    & 40   \\
	S II  & 0.4  & 7000 & 7.6 & 1.0 & 14   & 40   \\
	Ca II & 4000 & 7000 & 7.6 & 1.3 & 0    & 40   \\
	Ti II & 8.0  & 7000 & 7.6 & 1.0 & 0    & 40   \\
	Cr II & 3000 & 7000 & 7.6 & 1.0 & 0    & 40   \\
	Fe II & 50   & 7000 & 7.6 & 0.6 & 0    & 40   \\
	\enddata
\tablenotetext{*}{All velocities are given in $1000~\mathrm{km~s}^{-1}$}
\end{deluxetable}

\begin{deluxetable}{lcccccc}
	\tabletypesize{\scriptsize}
	\tablecaption{Fitting Parameters for 1999by Day 29 Post {\emph B} Maximum.
			\label{99byday029}}
	\tablewidth{0pt}
	\tablehead{
		\colhead{Element}         & \colhead{$\tau$}
		& \colhead{$T_{exc}$ (K)} & \colhead{$v_{phot}$} 
		& \colhead{$v_{e}$}    & \colhead{$v_{min}$}
		& \colhead{$v_{max}$} }
	
	\startdata
	H$V_{e}$~O I& 0.6  & 7000 & 7.6 & 1.6 & 0    & 40   \\
	O I   & 0.4  & 7000 & 7.6 & 1.0 & 14   & 16   \\
	Na I  & 3.0  & 7000 & 7.6 & 0.7 & 11   & 40   \\
	Mg II & 7.0  & 7000 & 7.6 & 1.0 & 12   & 13   \\
	Si II & 15   & 7000 & 7.6 & 0.5 & 0    & 40   \\
	S II  & 0.4  & 7000 & 7.6 & 1.0 & 14   & 40   \\
	Ca II & 4000 & 7000 & 7.6 & 1.3 & 0    & 40   \\
	Ti II & 8.0  & 7000 & 7.6 & 1.0 & 0    & 40   \\
	Cr II & 3000 & 7000 & 7.6 & 1.0 & 0    & 40   \\
	Fe II & 50   & 7000 & 7.6 & 0.6 & 0    & 40   \\
	\enddata
\tablenotetext{*}{All velocities are given in $1000~\mathrm{km~s}^{-1}$}
\end{deluxetable}
\clearpage

\clearpage\begin{deluxetable}{lcccccc}
	\tabletypesize{\scriptsize}
	\tablecaption{Fitting Parameters for 1999by Day 31 Post {\emph B} Maximum.
			\label{99byday031}}
	\tablewidth{0pt}
	\tablehead{
		\colhead{Element}         & \colhead{$\tau$}
		& \colhead{$T_{exc}$ (K)} & \colhead{$v_{phot}$} 
		& \colhead{$v_{e}$}    & \colhead{$v_{min}$}
		& \colhead{$v_{max}$} }
	
	\startdata
	H$V_{e}$~O I& 0.6  & 7000 & 7.5 & 1.6 & 0    & 40   \\
	O I   & 0.1  & 7000 & 7.5 & 1.0 & 0    & 40   \\
	Na I  & 3.5  & 7000 & 7.5 & 0.6 & 12   & 40   \\
	Mg II & 0.1  & 7000 & 7.5 & 1.0 & 12   & 13   \\
	Si II & 10   & 7000 & 7.5 & 0.6 & 0    & 40   \\
	S II  & 0.4  & 7000 & 7.5 & 1.0 & 14.5 & 40   \\
	Ca II & 4000 & 7000 & 7.5 & 1.3 & 0    & 40   \\
	Ti II & 7.0  & 7000 & 7.5 & 1.0 & 0    & 40   \\
	Cr II & 3300 & 7000 & 7.5 & 1.0 & 0    & 40   \\
	Fe II & 35   & 7000 & 7.5 & 0.8 & 0    & 40   \\
	\enddata
\tablenotetext{*}{All velocities are given in $1000~\mathrm{km~s}^{-1}$}
\end{deluxetable}

\begin{deluxetable}{lcccccc}
	\tabletypesize{\scriptsize}
	\tablecaption{Fitting Parameters for 1991bg Day 32 Post {\emph B} Maximum.
			\label{91bgday032}}
	\tablewidth{0pt}
	\tablehead{
		\colhead{Element}         & \colhead{$\tau$}
		& \colhead{$T_{exc}$ (K)} & \colhead{$v_{phot}$} 
		& \colhead{$v_{e}$}    & \colhead{$v_{min}$}
		& \colhead{$v_{max}$} }
	
	\startdata
	H$V_{e}$~O I& 0.6  & 7000 & 7.6 & 1.6 & 0    & 40   \\
	O I   & 0.5  & 7000 & 7.6 & 1.0 & 14   & 16   \\
	Na I  & 6.0  & 7000 & 7.6 & 0.6 & 11   & 40   \\
	Mg II & 4.0  & 7000 & 7.6 & 1.0 & 12   & 14   \\
	Si II & 10   & 7000 & 7.6 & 0.5 & 0    & 40   \\
	S II  & 0.1  & 7000 & 7.6 & 1.0 & 14.5 & 40   \\
	Ca II & 4000 & 7000 & 7.6 & 1.3 & 0    & 40   \\
	Ti II & 7.0  & 7000 & 7.6 & 1.0 & 0    & 40   \\
	Cr II & 3500 & 7000 & 7.6 & 1.0 & 0    & 40   \\
	Fe II & 40   & 7000 & 7.6 & 0.6 & 0    & 40   \\
	\enddata
\tablenotetext{*}{All velocities are given in $1000~\mathrm{km~s}^{-1}$}
\end{deluxetable}
\clearpage

\clearpage

\begin{figure}
	\begin{center}
	\includegraphics[scale=0.6]{f12}
	\caption{Si~II $\lambda$6355 velocities. Filled symbols 
		are CL~SNe~Ia and diamond symbols represent 
		velocities from our data. Other line velocity data comes 
		from \cite{turatto96, jha99, garnavich04} and 
		\cite{taubenberger08}.\label{si2vels}}
	\end{center}
\end{figure}
\clearpage{}

\clearpage{}

\end{document}